# Self-diffusivity as a function of density and temperature in crystalline solids and compensating rules for self-diffusion parameters in Carbon - Subgroup crystals

A. N. Papathanassiou, I. Sakellis and J. Grammatikakis

University of Athens, Physics Department, Solid State Physics Section, Panepistimiopolis, 15784 Zografos, Athens, Greece

## ABSTRACT

The self-diffusion coefficient of crystalline solids as a function of density and temperature may derive from thermodynamics concepts and an earlier elastic thermodynamic point defect model [P. Varotsos and K. Alexopoulos, Phys. Rev. B **15**, 4111 (1977); Phys. Rev. B **18**, 2683 (1978)]. Compensation laws ruling self-diffusion parameters in carbon-subgroup crystals obtained from theoretical calculations are predicted, as well.

Analytic functions that scale diffusivity in ultra-viscous liquids (which constitute an exceptional state of matter, characterizing 'solids that flow' [1]) with respect to density were derived recently [2, 3]. The scope was to understand why scaling of dynamic quantities scale with density and temperature in ultra-viscous liquids, which are characterized by a strong temperature dependence of the activation enthalpy (fragile liquids). These efforts were based on thermodynamical concepts and an earlier thermodynamic elastic model (the so-called cBΩ model [4]) that correlates the Gibbs energy for activation with the isothermal bulk modulus, combined with specific characteristics of the ultra-viscous state. Inspired by the above-mentioned approach, diffusivity vs density and temperature function for crystalline solids are derived in the present work. A pressure dependent diffusivity equation that applies to crystalline solids is modified through a generalized simple equation of state, the cBΩ model and proper temperature dependence for the Gibbs activation energy to a density and temperature function. Alternatively, we reach to the same point by combining the pressure dependent diffusivity equation solely with basic aspects of the cBΩ elastic model.

The mechanism of self-diffusion in crystals of the carbon subgroup C (diamond), Si, Ge α-Sn and Pb remains a matter of ongoing investigation [5]. These crystals have very large Debye temperature (i.e., for diamond, $\Theta_D$=2246K), making quantum effects appreciably important even at room temperature. Diffusion of point defects in the diamond family is important for both fundamental and applied research and technology. Self-diffusion calculations were recently reported in carbon-subgroup crystals by Magomedov [6]. An expression for the self-diffusion coefficient as a function of density ρ and temperature was proposed; i.e., $D(\rho, T) = D_d(\rho)\chi(\rho, T)$, where $D_d(\rho)$ is a function of the correlation factor, the inter-atomic spacing, the pack-density of atoms and the Debye temperature and χ(ρ,T) is the fraction of atoms having kinetic energy above a threshold value required to diffuse. In this paper, we shall show that diffusivity functions vs density and temperature mentioned in Ref. [6] result independently from fundamental thermodynamic concepts and the so-called cBΩ elastic point defect model [4], which asserts that the Gibbs activation energy is proportional to the isothermal bulk modulus and the mean atomic volume. The parameters of the diffusivity vs density and temperature equations are correlated with physical quantities of the crystal, which vary slightly with pressure; the latter reveals the scaling feature of the equation derived in this work (i.e., diffusivity isobars at different pressures collapse on a common (master) curve). Compensation laws relating self-diffusion entropy and enthalpy as well as self-diffusion entropy and volume observed in Ref. [6] are direct consequences of the cBΩ model.

Starting from the definition of the isothermal bulk modulus:

$$B \equiv -(\partial P/\partial \ln V)_T \tag{1}$$

and recalling that the density is $\rho \equiv m/V$, we get:

$$B = (\partial P/\partial \ln \rho)_T \tag{2}$$

To a first approximation, the bulk modulus can be described efficiently by a linear function (i.e., second order terms with respect to pressure are considered to be practically small):

$$B(P) = B_0 + (\partial B/\partial P)_T P \tag{3}$$

where $B_0$ denotes the zero (ambient) pressure value of the bulk modulus and $(\partial B/\partial P)_T$ is assumed to be (to a first approximation) roughly constant.

Eqs. (2) and (3) merge to:

$$\left(\frac{\partial P}{\partial \ln \rho}\right)_T = B_0 + (\partial B/\partial P)_T P \tag{4}$$

By integrating over pressure and density, we get the following equation of state (EOS):

$$\rho^{*(\partial B/\partial P)_T} = 1 + \frac{(\partial B/\partial P)_T}{B_0} P \tag{5}$$

where ρ* denotes furthermore the reduced density with respect with its value at ambient (zero) pressure. We note that Eq. (5) (which is employed to simplify the mathematical manipulation) is the so-called Murnaghan EOS and is based on the condition that B(P) is *practically* a linear function.

Self-diffusion in condensed matter can be regarded within the frame of the fluctuation of the volume when a 'flow event' occurs (e.g., migration of an atom from one equilibrium state to another one by passing over an (effective) saddle point). The activation volume controls the pressure evolution of the diffusivity $\upsilon^{act} \equiv (\partial g^{act}/\partial P)_T$, where $g^{act}$ denotes the Gibbs free energy for diffusion. Linear lnD(P) plots indicates $\upsilon^{act}$ is constant, while curved ones originate from the pressure dependence of $\upsilon^{act}(P)$ [4]. There is no physical reason to regard $\upsilon^{act}$ as constant; therefore, the compressibility of the activation volume is generally defined as $\kappa^{act} \equiv -(\partial \ln \upsilon^{act}/\partial P)_T$ [4], and can be positive, negative or zero.

A general equation for self-diffusion describing mono-vacancy mechanism for the three cubic Bravais lattices is:

$$D(P,T) = \lambda \alpha^2 \nu \exp(-g^{act}/kT) \qquad (6)$$

where D is the diffusion coefficient, λ is a geometrical factor, α is the inter-atomic spacing, ν is the vibrational frequency of the diffusing species (and related with the phonon frequency involved in the diffusion process) and k is the Boltzmann's constant. For a single mechanism of diffusion $g^{act} = g^m + g^f$, where $g^m$ and $g^f$ denote the free energy for migration and formation of vacancies, respectively; i.e., $g^{act}$ describes both the motion of carriers and the changes of their concentration induced by temperature and pressure variation. Differentiating Eq. (6) with respect to pressure and considering that pressure does not modify the geometrical factor λ, we get:

$$\left(\frac{\partial \ln D}{\partial P}\right)_T = -\frac{\upsilon^{act}(P)}{kT} + \frac{1}{B(P)}\left(\gamma_G - \frac{2}{3}\right) \qquad (7)$$

where $\gamma_G$ is the Grüneisen parameter and, as mentioned earlier, B(P) is linear. If $\upsilon^{act}$ is constant (i.e., $\kappa^{act}=0$), Eq. (7) yields a reduced diffusivity D* (which is the diffusivity D(P) reduced to the ambient (zero) pressure diffusivity value):

$$\ln D^*(P) = -\frac{\upsilon^{act}}{kT}P + \left(\gamma_G - \frac{2}{3}\right)\frac{1}{(\partial B/\partial P)_T}\ln\left(1 + \frac{(\partial B/\partial P)_T}{B_0}P\right) \quad (8a)$$

If $\upsilon^{act}(P)$ has a constant (positive) compressibility $\kappa^{act}$ (e.g., if the lnD(P) isotherms are curved upwards), we have $\upsilon^{act}(P) = \upsilon_0^{act}\exp(-\kappa^{act}P)$. For the case $\kappa^{act}P \ll 1$, the latter reduces to: $\upsilon^{act}(P) \cong \upsilon_0^{act}(1 - \kappa^{act}P)$. Thus, from Eq. (7), we get:

$$\ln D^*(P) = -\frac{\upsilon_0^{act}}{kT}P + \left(\frac{\upsilon_0^{act}\kappa^{act}}{2kT}\right)P^2 + \left(\gamma_G - \frac{2}{3}\right)\frac{1}{(\partial B/\partial P)_T}\ln\left(1 + \frac{(\partial B/\partial P)_T}{B_0}P\right) \quad (8b)$$

Alternatively, we may assume that $\kappa^{act}$ depends on pressure and consider that $1/\kappa^{act}(P)=B(P)$, i.e., the bulk modulus of the activation volume $B^{act}(P) \equiv 1/\kappa^{act}(P)$ has the same pressure dependence as that of the 'total' bulk modulus B(P): $1/\kappa^{act}(P) \equiv B^{act}(P) = B_0 + (\partial B/\partial P)_T P$. Within this condition, Eq. (7) reduces to:

$$\ln D^*(P) = -\frac{\upsilon_0^{act}B_0}{kT((\partial B/\partial P)_T - 1)}\left[\left(1 + \frac{(\partial B/\partial P)_T}{B_0}P\right)^{1-[(\partial B/\partial P)_T]^{-1}} - 1\right]$$
$$+ \left(\gamma_G - \frac{2}{3}\right)\frac{1}{(\partial B/\partial P)_T}\ln\left(1 + \frac{(\partial B/\partial P)_T}{B_0}P\right) \quad (8c)$$

We note that D*(P) is dimensionless and denotes the diffusivity reduced to its zero pressure value $D_0$. Pressure transforms to reduced density $\rho^*$ (i.e., the density reduced to its ambient pressure value) through Eq. (4). Eqs (8a), (8b) and (8c) can be rewritten respectively:

$$\ln D^*(\rho^*,T) = -\frac{\upsilon_0^{act} B_0}{kT(\partial B/\partial P)_T}(\rho^{*(\partial B/\partial P)_T}-1) + \left(\gamma_G - \frac{2}{3}\right)\frac{1}{(\partial B/\partial P)_T}\ln\rho^{*(\partial B/\partial P)_T}$$

(9a)

$$\ln D^*(\rho^*,T) = -\frac{\upsilon_0^{act} B_0}{kT(\partial B/\partial P)_T}(\rho^{*(\partial B/\partial P)_T}-1) + \left(\frac{\upsilon_0^{act} B_0}{2kT(\partial B/\partial P)_T^2}\right)(\rho^{*(\partial B/\partial P)_T}-1)^2$$
$$+\left(\gamma_G - \frac{2}{3}\right)\frac{1}{(\partial B/\partial P)_T}\ln\rho^{*(\partial B/\partial P)_T}$$

(9b)

and

$$\ln D^*(\rho^*,T) = -\frac{\upsilon_0^{act} B_0}{kT((\partial B/\partial P)_T-1)}\left[\rho^{*((\partial B/\partial P)_T-1)}-1\right]$$
$$+\left(\gamma_G - \frac{2}{3}\right)\frac{1}{(\partial B/\partial P)_T}\ln\rho^{*(\partial B/\partial P)_T}$$

(9c)

Whilst Eq. (9a) is based on the assertion that $\kappa^{act}=0$, Eq. (9b) derives from Eq. (8b) under the condition that $\kappa^{act}$ is of the order of the inverse of material's isothermal bulk modulus and was set roughly equal to $1/B_0$. Eq. (9c) stems from Eq. (8c) assuming that $1/\kappa^{act}(P) \equiv B^{act}(P) = B_0 + (\partial B/\partial P)_T P$.

Varotsos and Alexopoulos suggested that the bulk modulus is the elastic quantity that controls activation and established proportionality between activation Gibbs free energy and bulk modulus (cBΩ model) [7 8, 9, 10]. Research on key role of elastic models to understand the peculiar properties of viscous liquids was motivated by Dyre [1]. Experimental results for many different types of materials at various experimental conditions (pressure and temperature) support the validity of the cBΩ model. Thus, it seems that the bulk modulus manifests a migration process rather than shear modulus [11, 12]. According to the so-called cBΩ model [7, 8, 9, 10]:

$$g^{act} = cB\Omega$$

(10)

where c is practically constant [9] and Ω denotes the mean atomic volume. Note that the validity of Eq. (10) has been checked at ambient pressure in a wide range of solids extending from silver halides [13] to rare gas solids [14], in ionic crystals under gradually increasing uniaxial stress [15] in which electric signals are emitted before fracture (in a similar fashion as the electric signals detected before earthquakes [16, 17, 18, 19]), as well as in disordered polycrystalline materials [20]. Recently, diffusivity-density equations describing scaling of the dynamic properties of ultra-viscous liquids based on thermodynamic concepts and the cBΩ model were reported [2, 3]. Differentiating Eq. (10) with respect to pressure we get:

$$\upsilon^{act} = B^{-1}[(\partial B/\partial P)_T - 1]g^{act} \qquad (11)$$

The Gibbs activation energy is a decreasing function of temperature, in general. Since $g^{act} \equiv h^{act} - Ts^{act}$, where $h^{act}$ and $s^{act}$ denote the activation enthalpy and entropy respectively. $g^{act}(T) = f(T)T$, where f(T) is an adjustable function. Thermodynamics demand a significant increase of $h^{act}$ and $s^{act}$ with temperature. The excessive fall of $g^{act}$ is due to an increasing difference between $h^{act}$ and $s^{act}$ [21]. The latter aspect underlies the temperature variation of self-diffusivity parameters calculated for carbon sub-group crystals [6]. Subsequently, at zero pressure, Eq. (11) is rewritten as:

$$\frac{\upsilon_0^{act} B_0}{kT} \approx f(T)[(\partial B/\partial P)_T - 1] \qquad (12)$$

Introducing the latter relation into Eqs. (9a) - (9c) we get:

$$\ln D^*(\rho^*, T) = -\frac{f(T)[(\partial B/\partial P)_T - 1]}{(\partial B/\partial P)_T}(\rho^{*(\partial B/\partial P)_T} - 1) + \left(\gamma_G - \frac{2}{3}\right)\frac{1}{(\partial B/\partial P)_T}\ln \rho^{*(\partial B/\partial P)_T}$$
(13a)

$$\ln D^*(\rho^*, T) = -\frac{f(T)[(\partial B/\partial P)_T - 1]}{(\partial B/\partial P)_T}(\rho^{*(\partial B/\partial P)_T} - 1)$$
$$+ \left(\frac{f(T)[(\partial B/\partial P)_T - 1]}{2(\partial B/\partial P)_T^2}\right)(\rho^{*(\partial B/\partial P)_T} - 1)^2 + \left(\gamma_G - \frac{2}{3}\right)\frac{1}{(\partial B/\partial P)_T}\ln \rho^{*(\partial B/\partial P)_T}$$

(13b)

and

$$\ln D^*(\rho^*, T) = -f(T)\left[\rho^{*(\partial B/\partial P)_T - 1} - 1\right]$$
$$+ \left(\gamma_G - \frac{2}{3}\right)\frac{1}{(\partial B/\partial P)_T}\ln \rho^{*(\partial B/\partial P)_T} \quad (13c)$$

The above three equations correlate the (reduced) self-diffusion coefficient with (reduced) density and temperature (the dependence upon temperature is expressed through the function f(T), which is material dependent). Note that each one holds under certain conditions for the function $\kappa^{act}(P)$ (which actually describes how $\upsilon^{act}(P)$ changes upon pressure): i.e., $\kappa^{act}=0$, $\kappa^{act}(P)$=constant (and $\kappa^{act}P \ll 1$) and $\kappa^{act}(P)=1/B(P)$, where B(P) is the isothermal bulk modulus of the crystal, respectively. The parameters determining Eqs (13a)-(13c) are negligibly dependent on pressure. Thus, diffusivity isobars obtained at various pressure values collapse on a common (master) curve when expressed as a function of density and temperature. It seems that the aforementioned representation reveals a scaling behaviour of the diffusivity vs density and temperature, which is a novel idea in the field of transport in crystalline solids.

An alternative straightforward route toward diffusivity vs density and temperature function in crystalline solids through the cBΩ model is presented in the next: Eq. (7) (again, B(P) is roughly linear and the resulting EOS is that given by Eq. (5)) combined with a couple of the cBΩ formulas Eqs (10) and (11), yields:

$$\ln D^*(P) = -c\int_0^P \frac{[(\partial B/\partial P)_T - 1]\Omega(P)}{kT}dP + \left(\gamma_G - \frac{2}{3}\right)\frac{1}{(\partial B/\partial P)_T}\ln\left(1 + \frac{(\partial B/\partial P)_T}{B_0}P\right)$$

(14)

However, $\Omega = F_{APF} \rho_{at}^{-1}$, where $F_{APF}$ is the atomic packing factor (which is assumed to be constant) and $\rho_{at}$ denotes here the atomic density. Thus, Eq. (14) yields:

$$\ln D^*(\rho_{at}^*, T) = -\frac{cF_{APF}B_0}{kT}\left(\rho_{at}^{*\,(\partial B/\partial P)_T - 1} - 1\right) + \left(\gamma_G - \frac{2}{3}\right)\frac{1}{(\partial B/\partial P)_T}\ln \rho_{at}^{*\,(\partial B/\partial P)_T} \tag{15}$$

where $\rho_{at}^*$ denotes the atomic density reduced to its ambient (zero) pressure value. It is worth noticing that the latter derivation is based on the cBΩ model and the concept of the atomic packing fraction; neither the rpessure dependence of $\upsilon^{ac}$ nor the temperature dependence of $g^{act}$ (expressed through the f(T) function) are required.

The calculations of Magomedov [6] reveal two compensation laws:
(i)  The self diffusion entropy $s^{act}$ is proportional to the activation enthalpy $h^{act}$, and,
(ii) The self-diffusion entropy $s^{act}$ is linear with the activation volume $\upsilon^{act}$.

In this section, we shall prove that the cBΩ model underlies the above-mentioned proportionality rules. Within the frame of the cBΩ model $s^{act}$ and $h^{act}$ are interconnected through [22]:

$$\frac{s^{act}}{h^{act}} = -\frac{\beta B + (\partial B/\partial T)_P}{B - T\beta B + (\partial B/\partial T)_P} \equiv \Re \tag{16}$$

where β denotes the volume thermal expansion coefficient. The compensation rule (i) is actually that predicted by the cBΩ model (i.e., $s^{act} = \Re h^{act}$). Furthermore, by dividing a couple of equations derived from the cBΩ model [11] by differentiation of Eq. (10 in respect to temperature and pressure:

$$s^{act} = -c\Omega(\beta B + (\partial B/\partial T)_P)$$
$$\upsilon^{act} = c((\partial B/\partial P)_T - 1)\Omega$$

we get:

$$\frac{s^{act}}{\upsilon^{act}} = -\frac{\beta B + (\partial B/\partial T)_P}{(\partial B/\partial P)_T - 1} \equiv \Re' \tag{17}$$

The compensation law (ii) observed by Magomedov is that predicted by the cBΩ model through Eq. (17) (i.e., $s^{act} = \Re' \upsilon^{act}$). These laws, stemming from the cBΩ model, have been experimentally tested for diamond [12], lead (see Ref [4], pp. 99-194 and 275), white tin (see Ref [4], pp. 232-238 and 280) and may probably apply to the carbon subgroup.

We conclude that, based on fundamental thermodynamic concepts and the so-called cBΩ elastic point defect model, we derive (in different alternative routes) analytic equations of the self-diffusion coefficient as a function of density and temperature to describe self-diffusion phenomena in crystalline solids. The parametrs of these novel expressions for the diffusivity are negligibly depending on pressure indicating that the corresponding diffusivity isobars collected ay different pressures, scale on a master curve. Compensation rules interconnecting self-diffusion parameters in carbon-subgroup crystals stem from the cBΩ model.

# References

1. J. Dyre, Rev. Mod. Phys. **78**, 953 (2006)
2. A.N. Papathanassiou, Phys. Rev. E **79**, 032501 (2009)
3. A. N. Papathanassiou and I. Sakellis, J. Chem. Phys. **132**, 154503 (2010)
4. P.A. Varotsos and K.D. Alexopoulos (1986), *Thermodynamics of Point Defects and Their Relation with Bulk Properties*, Editors: S. Amelinckx, R. Gevers and J. Nihoul (North-Holland, Amsterdam) pp. 126-127
5. K. T. Koga and M.J. Walter, Phys. Rev. B **72**, 024108 (2005)
6. M.N. Magomedov, Semicoductors, **44**, 271 (2010)
7. P. Varotsos, Phys. Rev. B **13**, 938 (1976)
8. P. Varotsos, W. Ludwig and K. Alexopoulos, **18**, 2683 (1978)
9. P. Varotsos, J. Appl. Phys., **101**, 123503 (2007)
10. P.A. Varotsos and K.D. Alexopoulos (1986), *Thermodynamics of Point Defects and Their Relation with Bulk Properties*, Editors: S. Amelinckx, R. Gevers and J. Nihoul, pp. 269-306.
11. P. Varotsos and K. Alexopoulos, Phys. Stat. Solidi (b) **110**, 9 (1982)
12. P. Varotsos, Phys. Rev. B, **75**, 172107 (2007)
13. P. Varotsos and K. Alexopoulos, J. Phys. Chem. Solids **39**, 759 (1978)
14. P. Varotsos and K. Alexopoulos, Phys. Rev. B **30**, 7305 (1984)
15. P. Varotsos N. Sarlis and M. Lazaridou, Phys. Rev. B **59**, 24 (1999)
16. P. Varotsos, N.V. Sarlis, E. S. Skordas and M.S. Lazaridou, Phys. Rev. E **71**, 011110 (2005)
17. P.A. Varotsos, N.V. Sarlis, E. S. Skordas, H.K. Tanaka and M.S. Lazaridou, Phys. Rev. E **73**, 031114 (2006)
18. P.A. Varotsos, N.V. Sarlis, E. S. Skordas, H.K. Tanaka and M.S. Lazaridou, Phys. Rev. E **74**, 021123 (2006)
19. P. Varotsos, K. Eftaxias, M. Lazaridou, G. Antonopoulos, J. Makris and J. P. Poliyannakis, Geophys. Res. Lett. **23**, 1449 (1996)
20. P.A. Varotsos and K. Alexopoulos, Phys. Rev. B **24**, 904 (1981)
21. P. Varotsos and K. Alexopoulos, J. Phys. C: Solid State Phys. **12**, L761 (1979)
22. P.A. Varotsos and K.D. Alexopoulos (1986), *Thermodynamics of Point Defects and Their Relation with Bulk Properties*, Editors: S. Amelinckx, R. Gevers and J. Nihoul (North-Holland, Amsterdam) p. 163.